\documentclass[aps,prl,preprint,groupedaddress]{revtex4-1}
\usepackage{graphicx}
\begin{document}

%\preprint{}

\title{Ruling out dark matter interpretation of the galactic GeV excess by gamma-ray data of galaxy clusters}


\author{Man Ho Chan, Chung Hei Leung}
%\thanks{}
\affiliation{The Education University of Hong Kong}


\date{\today}

\begin{abstract}
Recently, some very tight constraints of annihilating dark matter have been obtained from gamma-ray data of the Milky Way and Milky Way dwarf spheroidal satellite galaxies. In this article, we report that there are two excellent galaxy clusters (A2877 and Fornax) which can provide interesting constraints for annihilating dark matter. The lower limits of the dark matter mass for the thermal relic annihilation cross section are 25 GeV, 6 GeV, 130 GeV and 100 GeV respectively for the $e^+e^-$, $\mu^+\mu^-$, $\tau^+\tau^-$ and $b\bar{b}$ channels. For some configuration of our working assumptions, our results improve the Fermi-LAT upper limits of annihilation cross sections by a factor of $1.3-1.8$ for wide ranges of dark matter mass for $e^+e^-$, $\mu^+\mu^-$ and $b\bar{b}$ channels, and a factor of $1.2-1.8$ for $\tau^+\tau^-$ channel with dark matter mass $\le 100$ GeV. These limits basically rule out most of the existing popular dark matter interpretation of the GeV excess in the Milky Way. 
\end{abstract}

%\keywords{}

\maketitle


\section{Introduction}
In the past few years, gamma-ray observations of the Milky Way initiated a hot research area about dark matter annihilation. Many studies \cite{Abazajian,Calore,Calore2,Daylan,Abazajian2} point out that the GeV excess in the Milky Way can be best explained by dark matter annihilation with dark matter mass $m \sim 10-100$ GeV via $\tau^+\tau^-$ or $b\bar{b}$ channels. However, later gamma-ray observations of the Milky Way dwarf spheroidal satellite (MW dSphs) galaxies by the Fermi-Large Area Telescope (Fermi-LAT) put strong constraints on the annihilation cross sections for different channels \cite{Ackermann,Albert}. These limits give a certain tension for the dark matter interpretation of the GeV excess in the Milky Way. Besides, some recent studies suggest that the GeV excess in the Milky Way can be explained by pulsars' emission \cite{Yuan,Petrovic,Gaggero,Brandt,Bartels2,Ploeg} or depletion of gamma rays in molecular clouds \cite{Boer}. Furthermore, recent results of Fermi-LAT gamma-ray detection strongly disfavour the dark matter interpretation of the GeV excess \cite{Ajello}. Therefore, the GeV excess in the Milky Way becomes a very controversial issue in dark matter astrophysics now.

Although most of the gamma-ray observations are now focusing on the Milky Way and the MW dSphs galaxies, some studies start to realize that galaxy clusters are also good candidates to constrain annihilating dark matter. For example, Ando \& Nagai (2012) \cite{Ando} find that the gamma-ray data from the Fornax cluster is quite good to constrain the parameters of dark matter annihilation because its cosmic ray emission is not very significant. The cross section constrained by using the Fornax cluster data is $<\sigma v> \le (2-3) \times 10^{-25}$ cm$^3$ s$^{-1}$ \cite{Ando}. Besides, Huang et al. (2012) \cite{Huang} use three years of Fermi-LAT data of several nearby galaxy clusters to constrain annihilating dark matter. The upper limits of annihilation cross section are also of the order $10^{-25}$ cm$^{-3}$ s$^{-1}$. Later studies of gamma-ray emission from galaxy clusters and galaxy groups give more stringent constraints on dark matter parameters. For example, Ackermann et al. (2015) \cite{Ackermann3} and Lisanti et al. (2017) \cite{Lisanti} exclude the thermal relic annihilating dark matter via $b\bar{b}$ channel for $m \le 40$ GeV and $m \le 30$ GeV respectively. Also, using new Fermi-LAT data of nearby galaxy clusters can provide stringent constraints for gamma-ray line feature of annihilating dark matter \cite{Liang}. Besides gamma-ray data, by using radio data of galaxy clusters, the annihilation cross sections for some channels can be constrained down to $\sim 10^{-26}$ cm$^3$ s$^{-1}$ \cite{Storm,Storm2,Beck}, though the uncertainties due to magnetic field strength are somewhat significant. Overall speaking, these limits are quite close to or even tighter than some of the parameters for the dark matter interpretation of the GeV excess in the Milky Way.

In this article, by using the latest gamma-ray data of galaxy clusters from Fermi-LAT, we show that the A2877 and Fornax clusters can give very stringent constraints for annihilating dark matter. Our results improve the current limits of annihilation cross sections for wide ranges of dark matter mass. These limits are tighter than the parameters involved for the dark matter interpretation of the GeV excess in the Milky Way.

\section{Method}
Generally speaking, there are two different kinds of gamma-ray emissions for annihilating dark matter in galaxy clusters. One is `direct emission' and the other is `indirect emission'. `Direct emission' means the gamma-ray photons produced when dark matter particles annihilate each other. Since dark matter annihilation also produces electron and positron pairs, these electron and positron pairs can interact with the thermal electrons in galaxy clusters via Bremsstrahlung cooling. This process would emit gamma-ray photons if the energy of the electrons or positrons is high enough. This kind of emission is regarded as `indirect emission'.

\subsection{Direct emission}
The total number of photons for direct emission with energy greater than $E_0$ produced by annihilating dark matter within a galaxy cluster can be calculated by
\begin{equation}
\dot{N}_{\rm dir}=4 \pi \int_0^{R_{200}} \rho_{DM}^2 <\sigma v>m^{-2}r^2dr \int_{E_0}^m \frac{dN_{\gamma}}{dE}dE,
\end{equation}
where $\rho_{DM}$ is the dark matter density profile, $R_{200}$ is the radius of a galaxy cluster, and $dN_{\gamma}/dE$ is the energy spectrum of gamma-ray produced per one annihilation, which can be obtained in \cite{Cirelli}. Based on the sample of galaxy clusters in \cite{Chen} and the observed limits of gamma-ray flux $\Phi_{\rm obs}$ obtained in \cite{Ackermann2}, there are two galaxy clusters (Fornax and A2877) which have the largest ratio $I/\Phi_{\rm obs}$, where $I=\int \rho_{DM}^2r^2dr$. Larger the ratio of $I/\Phi_{\rm obs}$ can give tighter constraints for annihilating dark matter. Therefore, we choose the Fornax and A2877 clusters to perform analysis. By assuming the dark matter density profiles of the A2877 and Fornax clusters follow the Navarro-Frenk-White (NFW) density profile $\rho_{DM}=\rho_sr_s^3[r(r+r_s)^2]^{-1}$ \cite{Navarro} and using the corresponding parameters in \cite{Anderson}, we can get the total photons produced with energy greater than $E_0$ for direct emission. 

On the other hand, recent numerical simulations show that substructures' contribution for dark matter annihilation is very significant. High dark matter density of the substructures can boost the annihilation signal to more than 10-100 times for galaxy clusters \cite{Gao,Anderhalden,Sanchez,Marchegiani}. However, the actual value of the boost factor in a galaxy cluster is controversial. Some high-resolution simulations can get the boost factor $B \sim 1000$ (e.g. Phoenix simulation project) \cite{Gao} while other studies point to a lower value of the order $B \sim 30$ \cite{Anderhalden,Sanchez}. Moreover, the effect of tidal interaction in galaxy clusters is also controversial. Some studies point out that tidal interaction would give a lower boost \cite{Sanchez} while some studies suggest that tidal stripping can increase the boost factor by $2-5$ times \cite{Bartels}. Here, for a conservative estimation, we neglect the effect of tidal interaction and take the lower bound of the boost factor, which can be described by a simple parametrization \cite{Sanchez}:
\begin{equation}
\log_{10}[B(M_{200})]= \sum_{i=0}^5b_i \times \left[ \ln \left( \frac{M_{200}}{M_{\odot}} \right) \right]^i,
\end{equation}
where $b_i=[-0.442, 0.0796, -0.0025, 4.77\times 10^{-6}, 4.77 \times 10^{-6}, -9.69\times 10^{-8}]$ and $M_{200}$ is the total mass of a galaxy cluster. The accuracy of this parametrization is better than 5\% in the mass range $10^6M_{\odot}<M_{200}<10^{16}M_{\odot}$ \cite{Sanchez}. For the A2877 and Fornax clusters, the total mass $M_{200}$ are $7.54\times 10^{14}M_{\odot}$ and $1.39 \times 10^{14}M_{\odot}$ respectively \cite{Anderson}. Therefore, the parametrization can be applied and the corresponding boost factors are 35 and 30 respectively. As discussed in \cite{Marchegiani}, the boost factor of the order $\sim 30-35$ are standard values for galaxy clusters. By taking these boost factors, we can calculate the direct emission flux for each galaxy cluster by $\Phi_{\rm dir}=\dot{N}_{\rm dir}/4 \pi D^2$, where $D$ is the distance of the galaxy cluster. Here, we are using the point-source approximation, which can be justified because we will compare the calculated fluxes with the `point-source upper limits' in \cite{Ackermann2}. 

\subsection{Indirect emission}
Dark matter annihilation also produces electron and positron pairs. These electrons and positrons would diffuse and cool down to a lower energy by synchrotron cooling, inverse Compton scattering, Bremsstrahlung cooling and Coulomb cooling. Since the diffusion coefficient for a typical galaxy cluster is about $D_0 \sim 10^{28}-10^{30}$ cm$^2$ s$^{-1}$ \cite{Colafrancesco}, the diffusion time scale for a 1 GeV electron is about $t_d \sim R_{200}^2/D_0 \sim 10^{17}-10^{19}$ s. However, the total cooling time scale for the above cooling processes is just $t_c \sim 10^{16}$ s. Therefore, the effect of diffusion is nearly negligible in galaxy clusters. The equilibrium electron energy spectrum can be simply obtained by \cite{Storm}
\begin{equation}
\frac{dN_e}{dE}=\int_0^{R_{200}} 4 \pi r^2 \rho_{DM}^2<\sigma v>m^{-2}[b(E)]^{-1}Y(E)dr,
\end{equation}
where $Y(E)=\int_E^m (dN'_e/dE')dE'$ is the integrated energy spectrum of electrons produced per one annihilation and $b(E)$ is the cooling function which is given by \cite{Colafrancesco}
\begin{equation} 
b(E)=b_{\rm IC}E_{\rm GeV}^2+b_{\rm syn}E_{\rm GeV}^2B_{\mu}^2+b_{\rm cou}n+b_{\rm brem}n,
\end{equation}
where $n$ is the thermal electron number density in cm$^{-3}$, $E_{\rm GeV}=E/1$ GeV, $B_{\mu}$ is the magnetic field strength in $\mu$G, $b_{\rm IC}=0.25 \times 10^{-16}$ GeV s$^{-1}$, $b_{\rm syn}=0.0254\times 10^{-16}$ GeV s$^{-1}$, $b_{\rm cou}=6.13 \times 10^{-16}[1+\log(\gamma/n)/75]$ GeV s$^{-1}$, $b_{\rm brem}=1.51 \times 10^{-16}[\log(\gamma/n)+0.36]$ GeV s$^{-1}$ and $\gamma$ is the Lorentz factor of a high-energy electron or positron. We follow the $\beta$-profile to model the thermal electron number density profile in a galaxy cluster:
\begin{equation}
n=n_0 \left(1+ \frac{r^2}{r_c^2} \right)^{-3\beta/2},
\end{equation}
where $n_0$, $r_c$ and $\beta$ are fitted parameters, which can be obtained in \cite{Chen}. For the magnetic field profile, we follow \cite{Storm} to use
\begin{equation}
B=B_0 \left[ \left(1+ \frac{r^2}{r_c^2} \right)^{-3\beta/2} \right]^{\eta},
\end{equation}
where we have assumed $B_0=4.7$ $\mu$G and $\eta=0.5$ for both galaxy clusters (the effect of the uncertainties in $B$ will be discussed below) \cite{Storm}.

For Bremsstrahlung emission (indirect emission), the total gamma-ray photons emitted per second for indirect emission with energy greater than $E_0$ can be given by \cite{Sarazin}
\begin{equation}
\dot{N}_{\rm ind}=\int_{E_0}^m d\epsilon \frac{d\sigma_{\rm brem}}{d \epsilon}nc \int_{E_0}^m \frac{dN_e}{dE}dE,
\end{equation}
where 
\begin{equation}
\frac{d\sigma_{\rm brem}}{d \epsilon}=\frac{32 \pi e^6}{3m_e^2c^5h \epsilon} \ln \left(\frac{2\gamma^2m_ec^2}{\epsilon} \right).
\end{equation}

By taking the same boost factors, we can calculate the indirect emission flux for each galaxy cluster by $\Phi_{\rm ind}=\dot{N}_{\rm ind}/4 \pi D^2$, with the point-source approximation.

\section{Result}
We can now obtain the total gamma-ray flux $\Phi=\Phi_{\rm dir}+\Phi_{\rm ind}$ emitted from each of the galaxy clusters. Based on the observational results from Fermi-LAT \cite{Ackermann2}, we obtain the fluxes for three lower limits of gamma-ray energy $E_0=0.5$ GeV, $E_0=1$ GeV and $E_0=10$ GeV. Also, we use the upper limits of the fluxes for these values of $E_0$ to constrain $<\sigma v>$ (see Table 1 for the point-source approximation upper limits for each $E_0$). 

\begin{table}
 \caption{The observed gamma-ray flux 95\% C.L. upper limits for the A2877 and Fornax clusters (point-source approximation) \cite{Ackermann2}. All the units for the gamma-ray fluxes are in $10^{-11}$ cm$^{-2}$ s$^{-1}$.}
\label{table1}
\begin{tabular}{@{}lcc}
   \hline
   & A2877 & Fornax \\
  \hline
Flux upper limits for $E \ge 0.5$ GeV & 8 & 11 \\
Flux upper limits for $E \ge 1$ GeV & 3.6 & 4.9 \\
Flux upper limits for $E \ge 10$ GeV & 0.19 & 0.26 \\
  \hline
 \end{tabular}
\end{table} 

In Fig.~1, we plot the upper limits for the annihilation cross section $<\sigma v>$ for 4 popular channels ($e^+e^-$, $\mu^+\mu^-$, $\tau^+\tau^-$ and $b\bar{b}$). We can see that the gamma-ray flux upper limits can give very stringent constraints for the annihilation cross section. In standard cosmology, the thermal relic annihilation cross section is $<\sigma v>=2.2 \times 10^{-26}$ cm$^3$ s$^{-1}$ \cite{Steigman}. The lower limits of dark matter mass for the thermal relic annihilation cross section are $25$ GeV, $6$ GeV, $130$ GeV and $100$ GeV for the $e^+e^-$, $\mu^+\mu^-$, $\tau^+\tau^-$ and $b\bar{b}$ channels respectively. Although these limits are nearly the same as the current one obtained by Fermi-LAT data of the MW dSphs \cite{Ackermann,Albert,Geringer}, overall speaking, we get tighter constraints for the upper limits of annihilation cross sections, except for the $\tau^+\tau^-$ channel with $m>100$ GeV. Our results improve the upper limits of annihilation cross sections by a factor of $1.3-1.8$ for wide ranges of $m$ for $e^+e^-$, $\mu^+\mu^-$ and $b\bar{b}$ channels, and a factor of $1.2-1.8$ for $\tau^+\tau^-$ channel ($m \le 100$ GeV) (see Fig.~2).

Furthermore, our results are more stringent than that obtained by other recent studies with the data of galaxy clusters. For example, recent studies using the data of Virgo cluster \cite{Ackermann3} and Coma cluster \cite{Marchegiani} obtain the lower limit of $m \sim 40$ GeV for dark matter annihilating via $b\bar{b}$ channel. Our results improve this limit by a factor of 2.5. Also, our results improve the limits obtained in \cite{Huang,Storm} (using the data of other clusters, such as Ophiuchus, A2199 and AWM7) by nearly an order of magnitude. Nevertheless, by using radio data of galaxy clusters, some recent studies can give more stringent constraints of annihilation cross sections. For example, Beck \& Colafrancesco \cite{Beck} obtain $m \ge 200$ GeV for the thermal relic annihilation cross section via $b\bar{b}$ or $\tau^+\tau^-$ by using the radio data of the Coma cluster. However, these studies constrain dark matter annihilation by synchrotron radiation which may suffer from large uncertainties of magnetic field strength. Although our study also includes the consideration of magnetic field strength, we consider the effect of Bremsstrahlung radiation due to magnetic field, but not synchrotron radiation. As discussed below, the uncertainties of magnetic field do not affect our result significantly. Therefore, using gamma-ray data is more reliable than using radio data for constraining annihilating dark matter, especially in using data of galaxy clusters. Based on the above discussion, we suggest that the Fornax and A2877 clusters are better candidates for constraining annihilating dark matter.

Moreover, our constraints here can safely rule out all of the existing popular dark matter interpretation of the GeV excess in our Milky Way (via $\tau^+\tau^-$ or $b\bar{b}$) \cite{Calore,Daylan,Abazajian2} (see Fig.~3). Besides, some recent models of dark matter interpretation of the positron excess and antiproton excess in our Milky Way are also ruled out (e.g. $\tau^+\tau^-$ and $b\bar{b}$ channels, see Fig.~3) \cite{Mauro,Cuoco}.

We have also checked with the effects by varying the magnetic field strength, dark matter profile and boost factor for both galaxy clusters. First of all, the magnetic field strength mainly affects the indirect emission for low energy ($E \le 10$ GeV). The contribution of indirect emission varies with dark matter mass mildly. The largest values of the ratio $\Phi_{\rm ind}/ \Phi_{\rm dir}$ for the $e^+e^-$, $\mu^+\mu^-$, $\tau^+\tau^-$ and $b\bar{b}$ channels are 26\%, 19\%, 3\% and 1\% respectively (mainly for $E \le 1$ GeV). In the above calculations, we used a typical central magnetic field strength $B_0=4.7$ $\mu$G and a typical index $\eta=0.5$. We have tried another extreme value $B_0=25$ $\mu$G \cite{Storm} and change the index $\eta$ to $0.4$, we find that the overall upper limits of annihilation cross sections are just larger by $<4$\%. Therefore, the effect of the systematic uncertainties of magnetic field strength (or contribution from indirect emission) is not very significant. The reason is that the most stringent constraints mainly come from the data of gamma-ray flux $E \ge 10$ GeV. Changing the magnetic field strength does not affect too much for the gamma-ray flux for $E \ge 10$ GeV.

For the dark matter profile, we have used the NFW profile to model the density profile of galaxy clusters. We have tried another profile, Einasto profile \cite{Merritt}, to model the Fornax and A2877 clusters as some recent studies suggest that the Einasto profile is a better description of dark matter halos on large scale \cite{Sereno}. We finally get tighter constraints after using the Einasto profile (see Fig.~4). Here, we only show the effect for the Fornax cluster because it gives the most stringent constraints. Generally speaking, using the NFW profile would give a more conservative result.

For the boost factor, as mentioned above, we used the most conservative values to model the boost factors of Fornax and A2877 clusters. These values depend on the virial mass of the clusters. By using an older value of virial mass for Fornax cluster ($M_{200}=1.01 \times 10^{14}M_{\odot}$) \cite{Sanchez2}, the boost factor of the Fornax cluster is 29, which gives only about 3\% larger for the upper limits. We have also tried another extreme empirical relation to calculate the boost factor (based on the simulation results from the Phoenix simulation project \cite{Gao}), the boost factor of the Fornax cluster is 525. The resultant constraints are much tighter (see Fig.~5).

Overall speaking, the assumed parameters and the profiles used are conservative. Therefore, the upper limits obtained are conservative limits.

\begin{figure}
\vskip5mm
 \includegraphics[width=120mm]{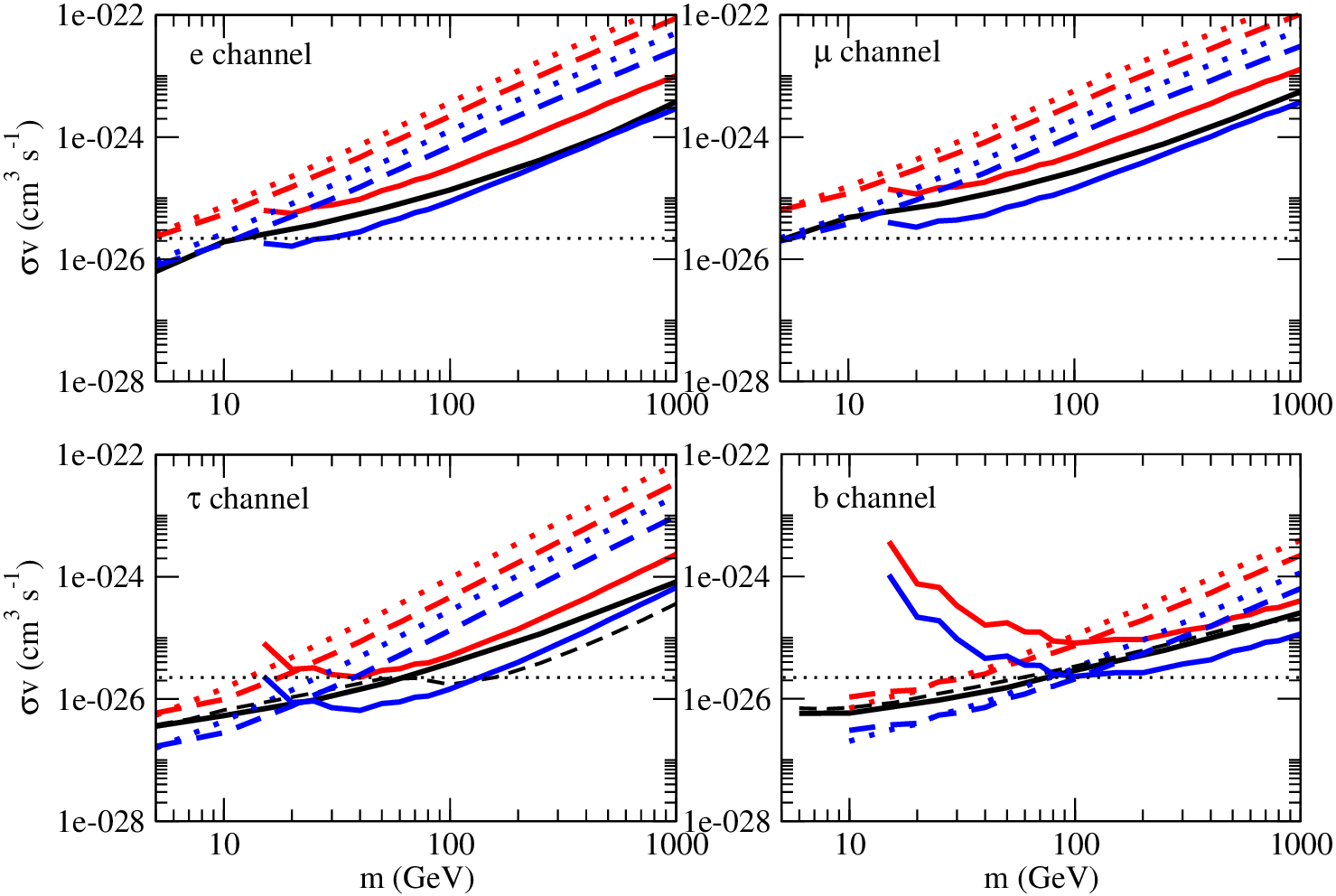}
 \caption{The upper limits of the annihilation cross section for 4 channels. The lines in red and blue represent the limits based on the data from A2877 and the Fornax cluster respectively (dotted line: $E_0=0.5$ GeV; dashed line: $E_0=1$ GeV; solid line: $E_0=10$ GeV). The black solid lines and dashed lines represent the upper limits of the annihilation cross section based on the Fermi-LAT data of the Milky Way dwarf spheroidal satellite galaxies \cite{Ackermann,Albert}. The black dotted lines represent the thermal relic cross section $<\sigma v>=2.2 \times 10^{-26}$ cm$^3$ s$^{-1}$ \cite{Steigman}.}
\vskip5mm
\end{figure}

\begin{figure}
\vskip5mm
 \includegraphics[width=120mm]{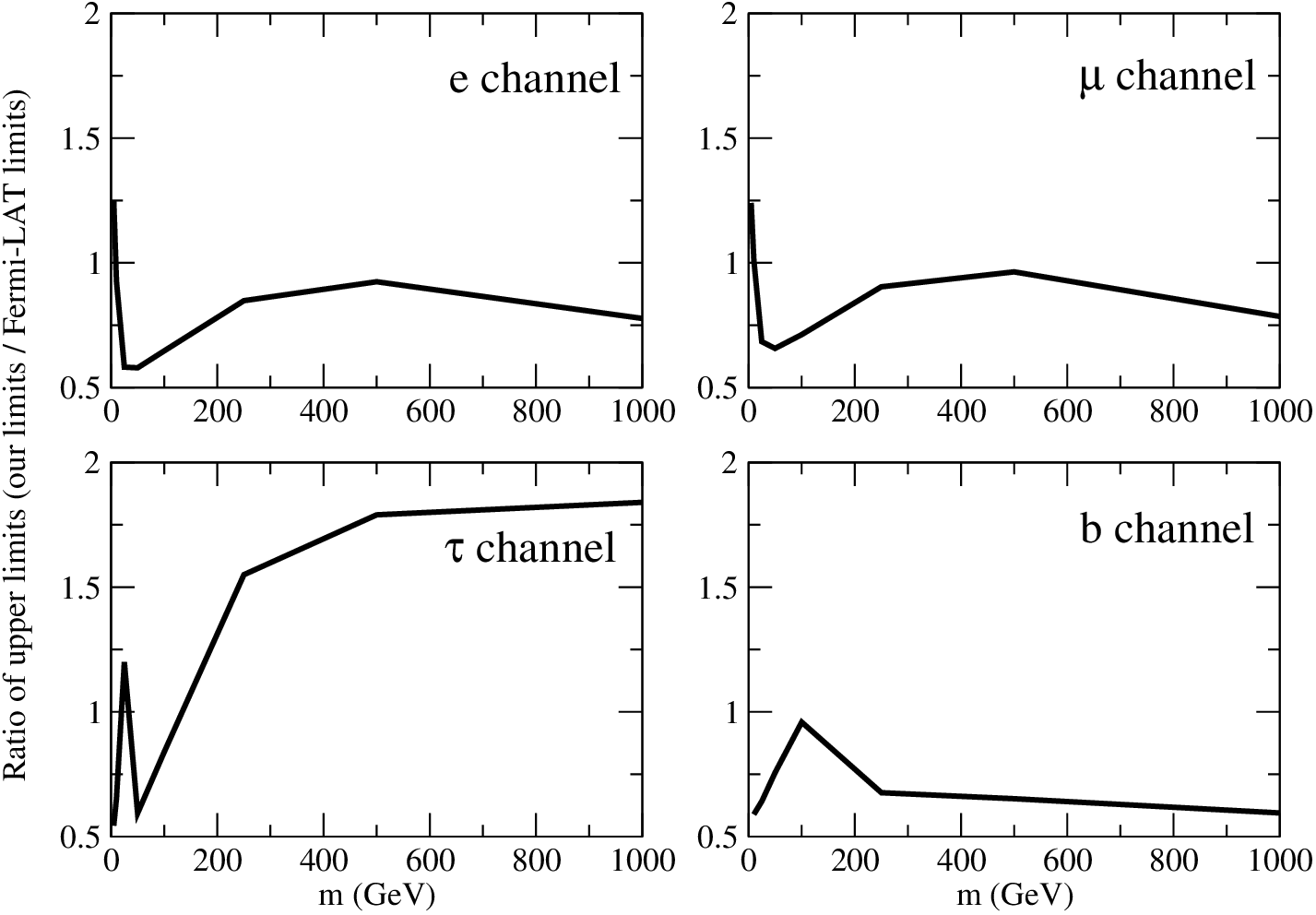}
 \caption{The ratio of the upper limits of the annihilation cross section obtained in this study to the Fermi-LAT upper limits in \cite{Ackermann,Albert} for 4 channels.}
\vskip5mm
\end{figure}

\begin{figure}
\vskip 10mm
 \includegraphics[width=120mm]{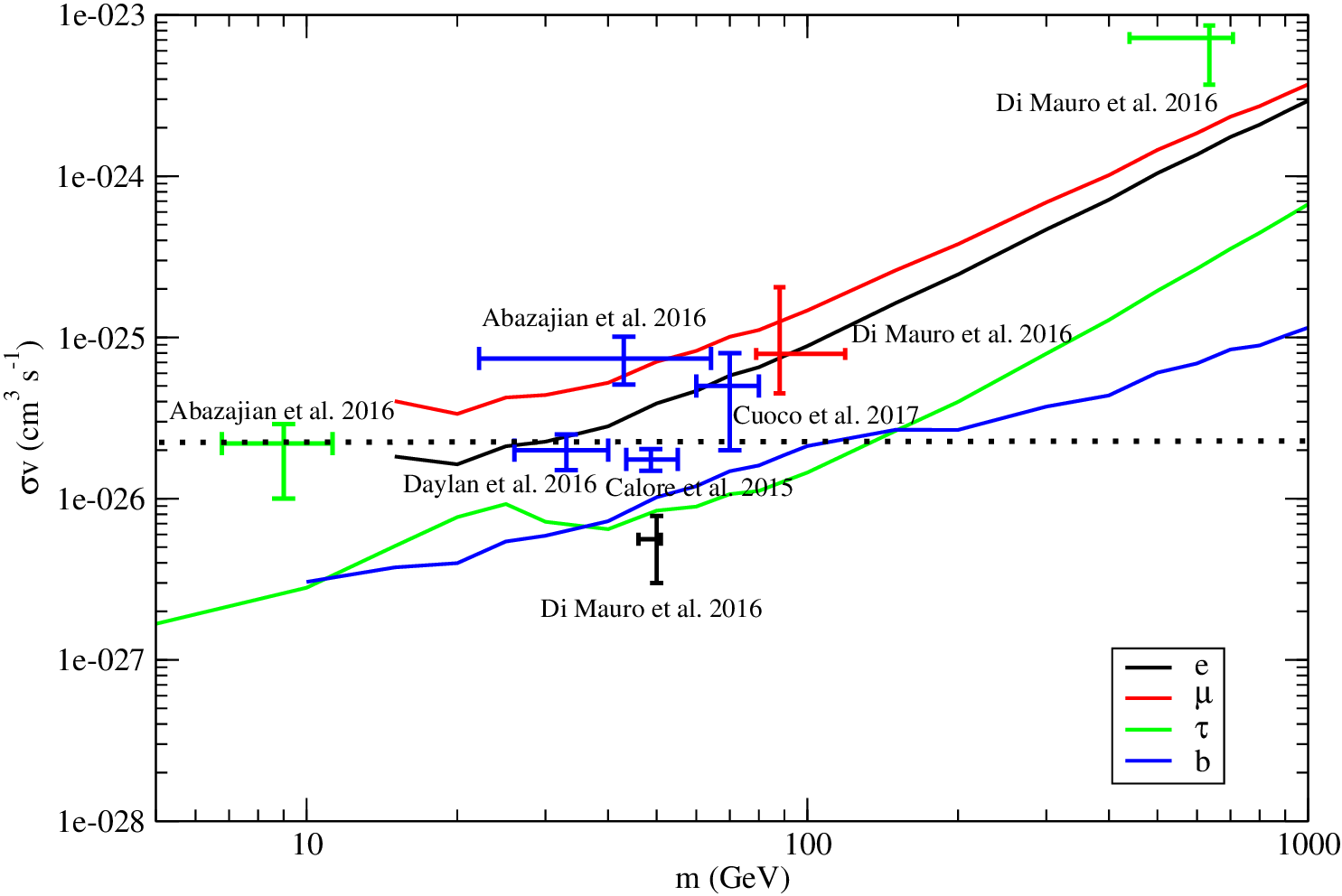}
 \caption{The solid lines represent the upper limits of the annihilation cross sections for the $e^+e^-$, $\mu^+\mu^-$, $\tau^+\tau^-$ and $b\bar{b}$ channels. The data points with 1$\sigma$ error bars are the results obtained in \cite{Calore,Daylan,Abazajian2,Mauro,Cuoco} for the dark matter interpretation of the GeV excess and positron excess. Here, different colors of solid lines and data points represent different annihilation channels. The dotted line is the thermal relic cross section.}
\vskip 10mm
\end{figure}

\begin{figure}
\vskip5mm
 \includegraphics[width=120mm]{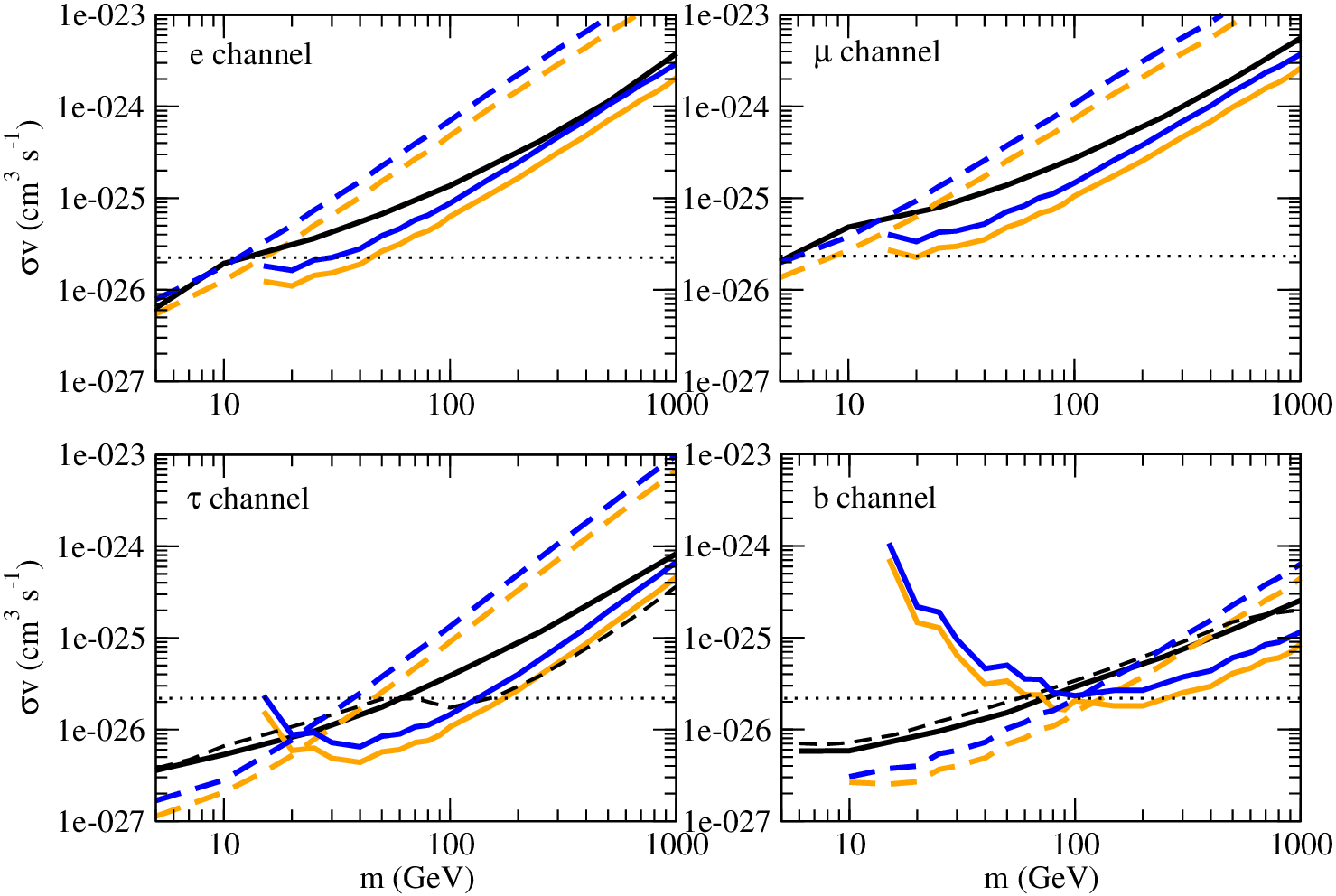}
 \caption{The upper limits of the annihilation cross section for 4 channels. The lines in blue and orange represent the limits using NFW and Einasto density profiles respectively (based on the data of the Fornax cluster only) (dashed line: $E_0=1$ GeV; solid line: $E_0=10$ GeV). The black solid lines and dashed lines represent the upper limits of the annihilation cross section based on the Fermi-LAT data of the Milky Way dwarf spheroidal satellite galaxies \cite{Ackermann,Albert}. The black dotted lines represent the thermal relic cross section $<\sigma v>=2.2 \times 10^{-26}$ cm$^3$ s$^{-1}$ \cite{Steigman}.}
\vskip5mm
\end{figure}

\begin{figure}
\vskip5mm
 \includegraphics[width=120mm]{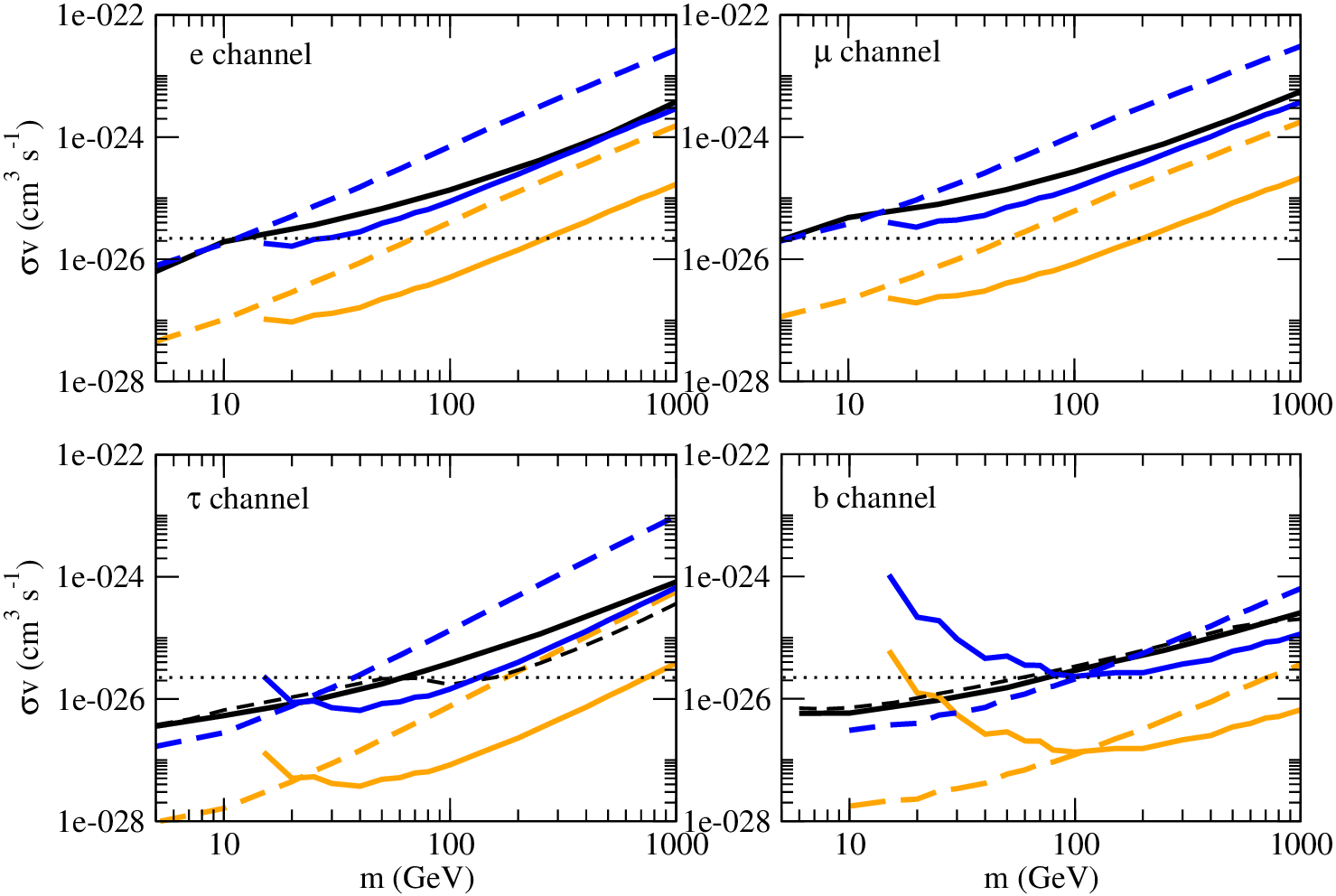}
 \caption{The upper limits of the annihilation cross section for 4 channels. The lines in blue and orange represent the limits using $B=30$ and $B=525$ respectively (based on the data of the Fornax cluster only) (dashed line: $E_0=1$ GeV; solid line: $E_0=10$ GeV). The black solid lines and dashed lines represent the upper limits of the annihilation cross section based on the Fermi-LAT data of the Milky Way dwarf spheroidal satellite galaxies \cite{Ackermann,Albert}. The black dotted lines represent the thermal relic cross section $<\sigma v>=2.2 \times 10^{-26}$ cm$^3$ s$^{-1}$ \cite{Steigman}.}
\vskip5mm
\end{figure}

\section{Discussion and Conclusion}
Recently, it has been reported that the excess gamma-ray emission in the Milky Way can be best explained by dark matter annihilation. The proposed dark matter mass is about $m \sim 10-100$ GeV. On the other hand, standard cosmology indicates that the thermal relic annihilation cross section should be $<\sigma v>=2.2 \times 10^{-26}$ cm$^3$ s$^{-1}$ so that it gives the correct amount of dark matter in our universe \cite{Steigman}. In this article, we obtain the upper limits of the annihilation cross sections for four popular channels by using the Fermi-LAT gamma-ray upper limits of two galaxy clusters (A2877 and Fornax). We improve the upper limits of annihilation cross sections by a factor of $1.3-1.8$ for wide ranges of $m$ for $e^+e^-$, $\mu^+\mu^-$ and $b\bar{b}$ channels, and a factor of $1.2-1.8$ for $\tau^+\tau^-$ channel (for $m \le 100$ GeV). Tighter constraints are obtained because we have considered the `indirect emission' (Bremsstrahlung emission) and the effect of substructures in galaxy clusters. In particular, the effect of substructures is significant in galaxy clusters as they are large and massive structures. Besides, we have used the most updated gamma-ray data for analysis, which provide more stringent limits for annihilating dark matter. Furthermore, among a large sample of nearby galaxy clusters, we found that the A2877 and Fornax clusters are the best candidates to constrain dark matter annihilation cross section because they have larger values of $I/\Phi_{\rm obs}$.

Besides, the lower limits of the dark matter mass for the thermal relic annihilation cross section are larger than the mass expected in dark matter interpretation of the gamma-ray excess. We show that most of the existing popular dark matter interpretation of the GeV excess in the Milky Way are ruled out. Our results basically support the latest Fermi-LAT analysis of the Milky Way GeV excess \cite{Ajello}.

In fact, searching dark matter particles is one of the most important tasks in particle physics. However, the null detection in direct detection experiments may indicate that dark matter mass is probably larger than 1 TeV \cite{Tan,Akerib}. Therefore, our results are quite consistent with these experimental results that dark matter mass may be much larger than the previous expected values. We expect that future observations of TeV gamma rays \cite{Abdallah,Archambault} and anti-proton detection \cite{Cavasonza,Feng} might provide some new insight of dark matter annihilation. 

\section{acknowledgements}
This work is supported by a grant from The Education University of Hong Kong (Activity code: 04234).

\section{Competing financial interests}
The authors declare no competing financial interests.

\section{Author contribution statement}
Design of the study and writing of the manuscript was performed by Man Ho Chan. Analysis of the results was performed by Man Ho Chan and Chung Hei Leung. All authors approved the submitted version of the manuscript.

\section{Data availability}
The datasets generated during and/or analysed during the current study are available from the corresponding author on reasonable request.

\end{document}